\documentclass[useAMS,usenatbib]{mn2e}

\input{psfig}                           %
\def\cygxu{\mbox{Cygnus X-1}}

\def\gro{\mbox{GRO J1655-40}}
\def\grs{\mbox{GRS 1915+105}}

\def\grsdstn{\mbox{GRS 1739-278}}

\def\gx{\mbox{GX 339-4}}

\def\novmus{\mbox{GRS 1121-68}}
\def\novoph{\mbox{GRS 1716-249}}

\def\cmmoinsdeux{\mbox{ cm}^{-2}}

\def\micrometre{\mbox{ } \mu \mbox{m}}

\def\kms{\mbox{ km\,s}^{-1}}

\def\kpc{\mbox{ kpc}}

\def\Msol{\mbox{ }M_{\odot}}

\def\mags{\mbox{ magnitudes}}

\def\ergpars{\mbox{ erg\,s}^{-1}}
\def\ergs{\mbox{ erg\,s}^{-1}}

\def\degp{{\rlap.}^{\circ}}

\def\asec{^{\prime \prime}}

\def\Av{A_{\rm v}}

\def\pom{\, \pm \,}

\def\nh{N_{\rm H}}

\def\ltsima{\; \buildrel < \over \sim \;}
\def\simlt{\lower.5ex\hbox{\ltsima}}            
\def\gtsima{\; \buildrel > \over \sim \;}
\def\simgt{\lower.5ex\hbox{\gtsima}}            

\begin{document}
\title[NIR Observations of Black Hole Candidates]
{Near Infrared Observations of Galactic Black Hole Candidates\thanks{Based on observations collected at the European Southern
Observatory, La Silla, Chile (proposals: 51.6-0018, 53.6-0015, 59.D-0719 \& 61.D-0542).}}
\author[S. Chaty et al.]
{S.~Chaty $^{1,2}$, I.F.~Mirabel $^{2,3}$, P.~Goldoni $^2$,
S.~Mereghetti $^4$, P.-A. Duc $^2$, J.~Mart\'{\i} $^5$, 
\newauthor R.P.~Mignani $^6$ \\
$^1$ Department of Physics and Astronomy, The Open University, Walton Hall,
Milton Keynes, MK7 6AA, 
United Kingdom\thanks{S.Chaty@open.ac.uk}\\
$^2$ Service d'Astrophysique, DSM/DAPNIA/SAp, CEA/Saclay,
L'Orme des Merisiers, B\^at. 709, F-91 191 Gif-sur-Yvette, Cedex, France \\
$^3$ Instituto de Astronom\'{\i}a y F\'{\i}sica del Espacio C.C. 67, 
Suc. 28. 1428, Buenos Aires, Argentina \\
$^4$ Istituto di Fisica Cosmica ``G.Occhialini'',
via Bassini 15, I-20133 Milano, Italy \\
$^5$ Departamento de F\'{\i}sica, Escuela Polit\'ecnica Superior,
Universidad de Ja\'en, Calle Virgen de la Cabeza, 2, E-23071 Ja\'en, Spain \\
$^6$ ESO, Karl-Schwarzschild Strasse 2, D-85748 Garching-bei-M\"unchen, Germany
}
\date{Received / Accepted} 
\maketitle

\begin{abstract}
We report on several 
European Southern Observatory (ESO) 
near-infrared (NIR) observational campaigns aimed 
at understanding the nature of 
galactic black hole candidates.
Our results, including NIR photometry of the sources
$\gro$, $\grsdstn$, $\novoph$, $\novmus$ and $\gx$, 
show that all the sources but $\gro$ are consistent with low-mass
stars as the companion star of the binary system.

By locating the counterparts on a colour-magnitude diagram (CMD),
we better constrain the spectral type of the companion star of three
of the systems considered here, and confirm a fourth one.
The spectral types are respectively: 
M0-5 V for $\novoph$, 
F8-G2 III for $\gx$
and later than F5 V for $\grsdstn$.
We confirm the already known spectral type of the companion
in $\novmus$ (K0-5 V).
The location of $\gro$ on the CMD 
is consistent with the sub-giant luminosity
class and with this source crossing the Hertzsprung gap.
However, a non-stellar emission seems to contribute to the NIR
flux of this source.

\end{abstract}


\section{Introduction}

Optical follow-up observations of transient X-ray sources
are fundamental to fully understand the nature of these
systems and characterize the accretion. 
Most important, the measurement of their mass function
through optical observations performed in quiescence
is the best way to confirm the presence of a black hole
(see e.g. \citealt{charles:1999}).
Most of the bright transient X-ray sources discovered in recent years
are located in the Galactic Bulge region.
Owing to the strong absorption present along this direction of the Galaxy,
NIR observations are one of the best ways of constraining 
the properties of these systems.
Furthermore, the
disk emission is much weaker in the NIR band allowing
a much better probe of the nature of the companion star.
One approach is to perform spectroscopic observations 
(see e.g. \citealt{bandyopadhyay:1999}), but for faint sources
or for the sources lacking prominent emission lines, 
photometric observations are more efficient for deriving
the NIR spectral energy distribution of the source 
(see e.g. \citealt{chaty:1996a}).

We observed the infrared counterparts of several galactic bulge
hard X-ray sources, in order to better constrain the spectral type of the 
mass donor star.
Here we report on our results on five galactic black hole
candidates: $\gro$, $\grsdstn$, $\novoph$, $\novmus$ and $\gx$.
Since all the sources we observed show variations, 
we concentrate on the magnitudes 
obtained around the emission minimum of our observations.
Assuming that all the emission at this time came from
the photosphere of the companion star, we can compare our
data with infrared magnitudes of different stellar spectral
types. 
Any anomalous colour distribution can then be related to the
source spectral state, known from high energy observations. 
Obviously, this method has limitations, since the infrared flux
may be contaminated. 
There are a number of possible sources of contamination: 
emission from the accretion disk; X-ray heating of the secondary;
the presence of ejected material,
as was the case during flares of the superluminal source
$\grs$ \citep{mirabel:1998a}; the presence of surrounding
dust or of an extended atmosphere. 
Nonetheless, since the emission provides at least an upper limit
to the source flux, this constrains the nature of the companion star,
and in addition can also give some information regarding
 sources of contamination.

In Section \ref{observations} we describe our method,
and then go through each source in turn, summarizing
the current knowledge and describing our observations 
and results. Discussion of the results and conclusions 
are in Section \ref{discussion}.
Some of the results presented here were partly reported in 
\citet{chaty:1998}. 

\section{Observations and results} \label{observations}

The general parameters of these sources, including the distance and 
the absorption that will be used in this paper, are given in Table
\ref{table_sources}.
In Table \ref{table_obs} we give a log of the observations 
that were mostly obtained at the ESO/MPI (Max Planck Institute) 
2.2~m telescope in La Silla (Chile) using the IRAC2b camera.
The IRAC2b camera, mounted at the F/35 infrared adapter of the telescope,
is a Rockwell 256$\times$256 pixels Hg:Cd:Te NICMOS 3 
infrared array detector. 
It was used with the lens C, providing an image scale of 
$0.49 \mbox{ arcsec/pixel}$ 
and a field of $136 \times 136 \mbox{ arcsec}^{2}$. 
The typical seeing for these observations was $1.2 \mbox{ arcsec}$.

Each final image is the result of the median-filtering 
of at least 9 frames of 1 minute exposure each 
(depending on the observations).
An image of the sky was taken after each image of the object
(offset by $30 \asec$),  
to allow subtraction of the infrared sky emission. 
The images were further treated by removal of the dark current 
and correction with a dome flat field, 
and we carried out absolute photometry by calibration
obtained with the observation of different standard stars.
This work was performed with the IRAF procedures, using the DAOPHOT package for   photometry in crowded fields.


\begin{table*}
\begin{flushleft}
\begin{tabular}{clllllllll} \hline
{\rm Source} & {\rm gal. coord.} & {\rm L$_{2-10keV}$} & {\rm L$_{opt}$} 
& {\rm L$_{6cm}$} & {\rm Period} & {\rm Distance} & {\rm $\nh$ } & {\rm A$_V$} \\ 
            &                & $\ergs$       & ($m_V$ mag)& (mJy)    &       &
 (kpc)        & (10$^{22}$cm$^{-2}$) & (mag) \\ \hline


GRO J1655-40 &
$l^{II} = 344 \degp 98$ 
& $2 \times 10^{37}$ & 
[17.4-16.8] & [0-2500] & $2.62157$d & 3.0      & [0.3/0.8]      & 2.46 \\
Nova Scorpii 94 & $b^{II} = +2 \degp 46$ & & &  & $\pm 13$ s & & 0.44 \\
& bai95 & gre96a & oro97 &       hje95 & oro97 &   gre96a &   nag94
\\ \hline


GRS 1739-278  &
$l^{II} = 0 \degp 66$ & 
$5 \times 10^{37}$ & 
23.2 & [1.1-4.7] & - & 8.5 &[$1.1/4.8]$& 11.18 \\
& $b^{II} = +1 \degp 17$ &
& $\pm 0.3$ & & &          & $2.0$ \\
& mar97 & gre96b & mar97 &       hje96 & & mar97 & gre96b/bor96
\\ \hline


GRS 1716-249 &
$l^{II} = 0 \degp 20$ & 
$2 \times 10^{38}$ & 
16.65 & [0.5-4.4] & $\sim 14.7$h & [2/2.8]    & 0.4             & 2.24 \\
Nova Ophiuchi 93 & $b^{II} = +6 \degp 99$ &
& & $\pm 0.1$ & & 2.4 \\
& del94 & del94 & del94 &       del94 & mas96 & del94 & tan93 
\\ \hline


GRS 1121-68 &
$l^{II} = 295 \degp 30$
& $10^{37}$ & 
[20.35-13.3] & [3-137] & 10.5h & [1.4/4]    & 0.22           & 1.23 \\
Nova Muscae 91 & $b^{II}=-7 \degp 07$ & & $\pm 0.05$ & & & 2.8  \\
& del 91 & del 91 & mcc92/del91       & kes91 & bai92 & del91/sha97 & gre94
\\ \hline


GX 339-4  &
$l^{II}=338 \degp 94$
& $2 \times 10^{36}$ & 
[21-15] & [0-6.5] & 14.8h & [5.6/7]    & 0.6          & 3.72 \\
4U 1658-48 & $b^{II}=-4 \degp 33$ & & &  & & 5.6 \\
& dox79 & rub98 & dox79/gri79       & fen97 & cal92 & sha01 
& pre91 & zdz98
\\ \hline

\end{tabular}
\end{flushleft}
\caption[]{\label{table_sources} {Parameters of the sources:
    coordinates, fluxes in different energy domains, 
orbital period when known, distance, hydrogen column density N$_H$ 
and absorption A$_V$. 
The variations, if any, are given by {\it [min-max]}. 
The error is given in the following line. 
The reader should refer to Table \ref{table_obs} to see the
NIR variations of these sources.
Concerning the distance and column density, we reported the interval 
in which they are constrained, including the error, by {\it [min/max]},
and the value choosen in this paper is given in the following line.
}
The references are:
  bai92:  \citealt{bailyn:1992}, 
  bai95:  \citealt{bailyn:1995}, 
  bor96: \citealt{borozdin:1996},
  cal92:  \citealt{callanan:1992}, 
  del91:  \citealt{dellavalle:1991}, 
  del94:  \citealt{dellavalle:1994}, 
  dox79:  \citealt{doxsey:1979},
  fen97:  \citealt{fender:1997c},
  gre94: \citealt{greiner:1994},
  gre96a: \citealt{greiner:1996a},
  gre96b: \citealt{greiner:1996b},
  gri79:  \citealt{grindlay:1979},
  hje95:  \citealt{hjellming:1995a},
  hje96:  \citealt{hjellming:1996},
  kes91:  \citealt{kesteven:1991},
  mar97:  \citealt{marti:1997a},
  mas96:  \citealt{masetti:1996}, 
  mcc92:  \citealt{mcclintock:1992},
  nag94:  \citealt{nagase:1994}, 
  oro97:  \citealt{orosz:1997},
  pre91:  \citealt{predehl:1991},
  rub98:  \citealt{rubin:1998},
  sha97: \citealt{shahbaz:1997},
  sha01: \citealt{shahbaz:2001},
  tan93: \citealt{tanaka:1993},
  zdz98: \citealt{zdziarski:1998}.
}
\end{table*}
\begin{table*}
\begin{flushleft}
\begin{tabular}{llllllll} 
\hline \noalign{\smallskip}
{\rm Source} & {\rm Date} & {\rm JD} & {\rm Tel} & {\rm ref} & {\rm $J$}($1.25 \micrometre$) & {\rm $H$}($1.65 \micrometre$) & {\rm $K$}($2.2 \micrometre$)\\
\noalign{\smallskip}
\hline \noalign{\smallskip}
$\gro$       & 19/07/97 & 2\,450\,648 & 2.2 &     & $13.43 \pm 0.03$ & $12.87 \pm 0.02$ & $12.07 \pm 0.01$ \\
& 07-10/99 & 2\,451\,421 & CTIO & gre01 & $13.7-14$ & - & $13.1-13.4$ \\ 
\hline
%
$\grsdstn$   & 01/04/96 & 2\,450\,174 & 2.2 & mar97 & $16.3 \pm 0.1$ & $15.5 \pm 0.2$   & $14.9 \pm 0.1$ \\
& 20/07/97 & 2\,450\,649 & 2.2 &       & $18.7 \pm 0.6$ & -                & $15.4 \pm 0.25$ \\
& 07/07/98 & 2\,451\,001 & 2.2 &       & $\geq 19.2$    & -                & $\geq 18.4$ \\  \hline
%
$\novoph$    & 08/07/94 & 2\,449\,539 & 2.2 &       & $16.2 \pom 0.3$ &              - & $15.5 \pom 0.3$ \\
& 19/07/97 & 2\,450\,648 & 2.2 &       & $19.4 \pom 1.2$ & $19.2 \pom 1$  & $18.3 \pom 1$ \\
& 06/07/98 & 2\,451\,000 & 2.2 &       & $\geq 19.38$    & $\geq 18.72$   & $\geq 17.77$ \\ \hline
%
$\novmus$    
& 07/07/94 & 2\,449\,540 & 2.2 &       & $17.5 \pm 0.25$ & -             & $16.7 \pm 0.25$ \\
& 18/02/95 & 2\,449\,766& AAT& sha97 & -              & $17.12 \pm 0.12$ & -               \\
& 19/07/97 & 2\,450\,648 & 2.2 &       & -               & -             & $17.23 \pm 0.25$ \\
& 06/07/98 & 2\,451\,000 & 2.2 &       & $\geq 17.5$  & $17.2 \pm 0.5$ & $\geq 16.5$ \\ \hline
%
%
$\gx$        & 03/06/93 & 2\,449\,141 & 2.2 &       & -             & -                  & $15.2 \pm 0.3$ \\
& 04/06/93 & 2\,449\,142 & 2.2 &       & $16.2 \pm 0.3$ & -                 & -              \\
& 06/07/94 & 2\,449\,539 & 2.2 &       & $15.3 \pm 0.1$ & $14.9 \pm 0.1$ & $14.5 \pm 0.1$ \\
& 08/07/94 & 2\,449\,541 & 2.2 &       & $15.4 \pm 0.1$ & -               & $14.4 \pm 0.1$ \\
& 19/07/97 & 2\,450\,648 & 2.2 &      & $14.21 \pm 0.03$ & $13.13 \pm 0.01$ & $12.41 \pm 0.01$\\
\noalign{\smallskip}
\hline
\end{tabular}
\end{flushleft}
\caption[]{\label{table_obs} {\rm Infrared observations of the sources}.
The exposure times were respectively 10, 10, 15 and 9 minutes 
in 1993, 1994, 1997 and 1998.
The two magnitudes quoted for $\gro$ in 1999 are the interval of the
ellipsoidal variations. \\
gre01: \citealt{greene:2001};
mar97: \citealt{marti:1997a};
sha97: \citealt{shahbaz:1997} \\
2.2: 2.2~m La Silla Telescope (ESO, Chile), with the instrument IRAC2b. \\
AAT: 3.9~m Anglo-Australian Telescope (AAT), with the infrared array IRIS. \\
CITO: Cerro Tololo Inter-American Observatory, with ANDICAM on the 1.0 m
Yale telescope
}
\end{table*}

\begin{figure*}
\centerline{\psfig{file=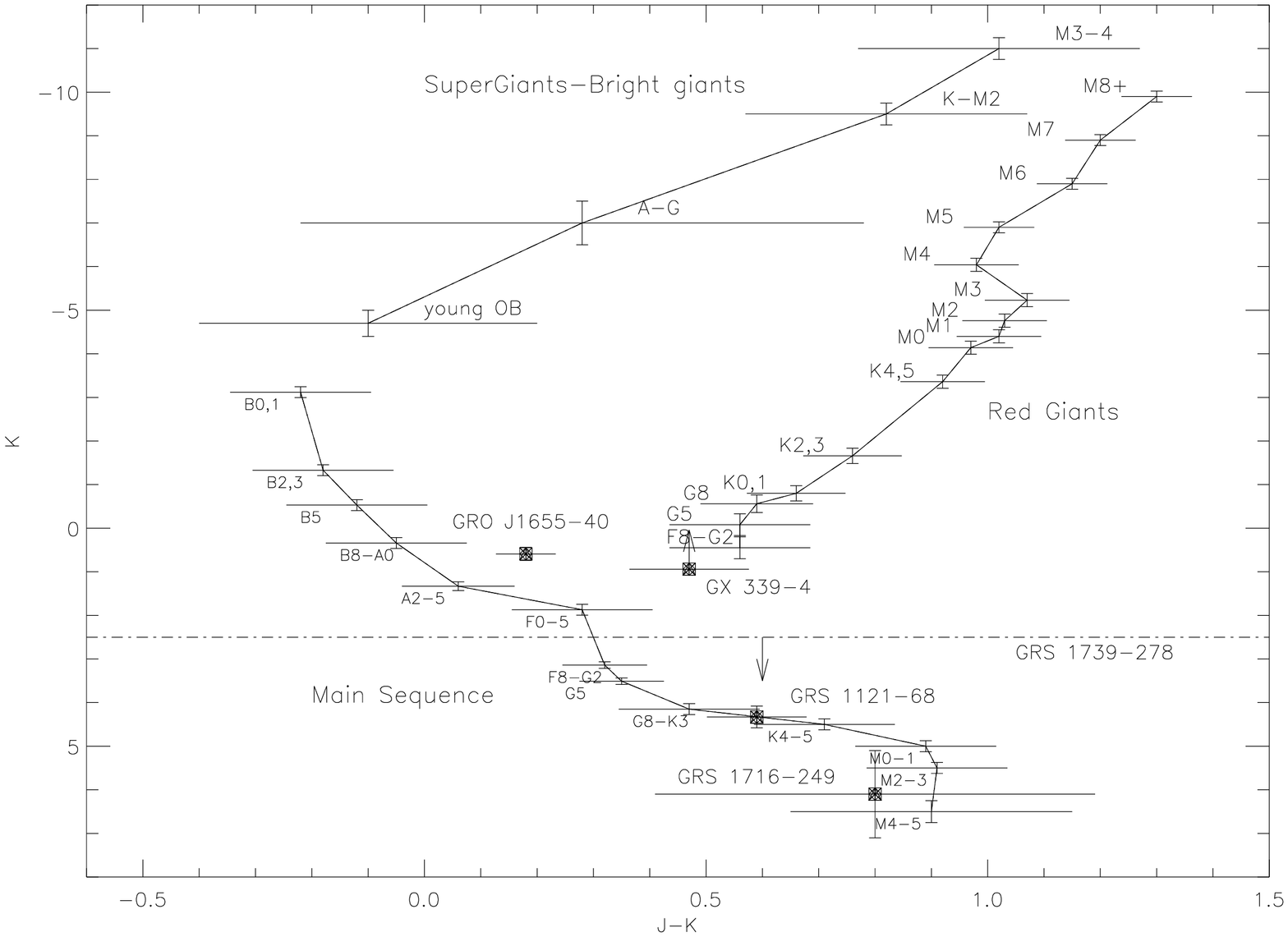,angle=90,width=18.cm}}
\caption[]{\label{hr} Near-infrared colour-magnitude diagram
of galactic stars \citep{ruelas-mayorga:1991}, 
with the superimposed counterparts of $\gro$, $\gx$, $\novoph$
and $\novmus$.
The error bars on $\gro$ are due to the ellipsoidal variations,
and the position of $\gx$ corresponds to the lower limit of the distance.
The dash-dot line is the upper limit for $\grsdstn$ 
(see section \ref{grsdstn_obs}).
This diagram shows that except for $\gro$, which is consistent
with an intermediate mass system, all sources are
consistent with low mass systems.}
\end{figure*}
\begin{figure*}
\centerline{\psfig{file=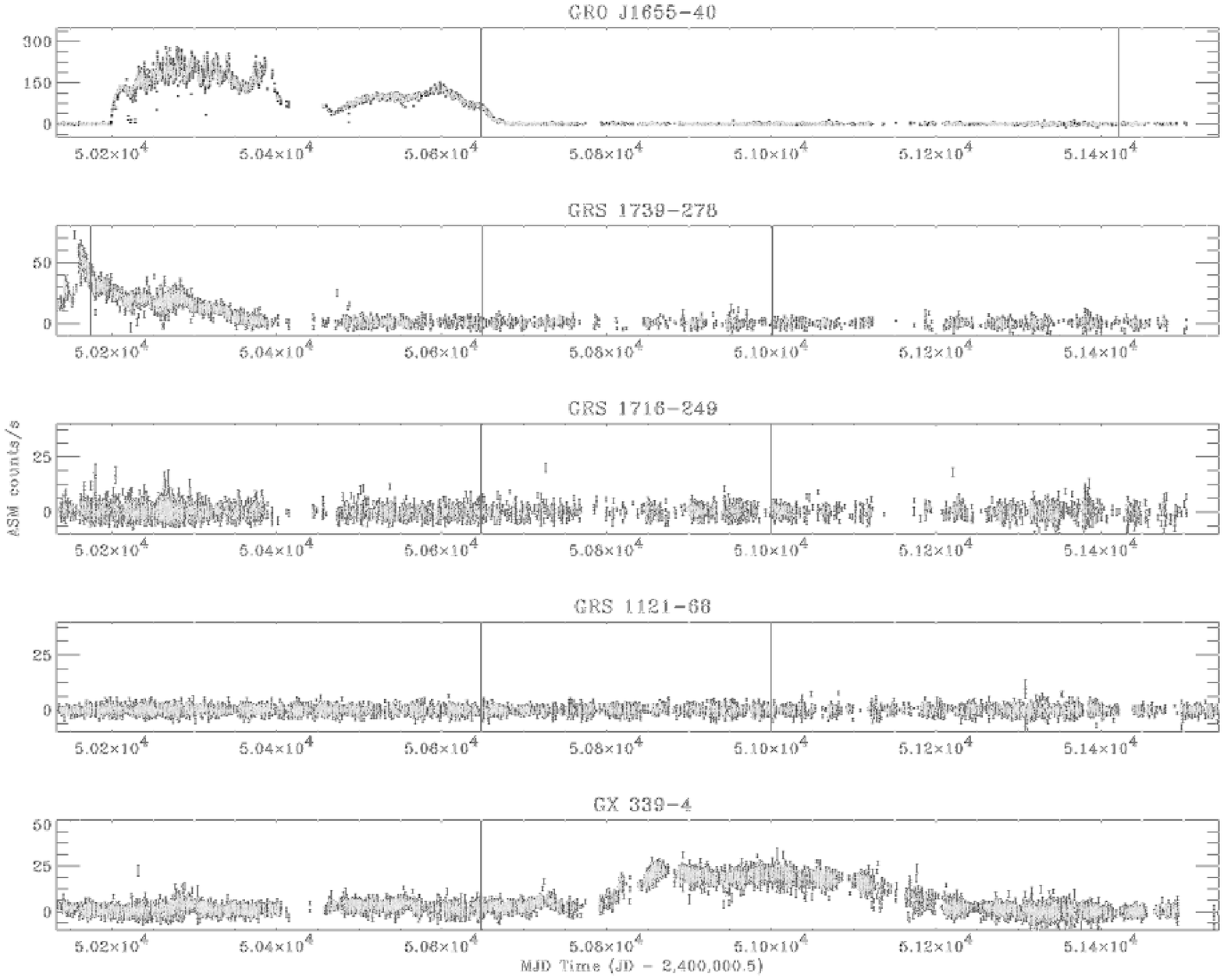,angle=90,width=18.cm}}
\caption[]{\label{asm} 
{\it RXTE/ASM} lightcurves of $\gro$, $\grsdstn$, $\novoph$, $\novmus$ and 
$\gx$ from 1996, February 20th
to 1999, December 31st. These quick-look results
show the observations 
in the high-energy band 2-12 keV. 
The time of observations are indicated by the vertical lines. 
For comparison, the Crab nebula flux is about 75 ASM counts/s.
Concerning the uncertainties, 
the light curve of the Crab nebula gives a measure of the rms error of 5\%.
However, the error bars are an underestimate of the error, if sources present
different spectral shape than the Crab, are located in crowded regions,
and/or for different reasons.
Therefore, we point out that, although we show here all the sources
for completeness, there is no flux detected by the {\it ASM} 
from the sources $\novoph$ and $\novmus$ during the period 
of observation shown here.
}
\end{figure*}

To estimate the nature of the companion stars,  
we compare their NIR absolute magnitudes with those of template 
stars, using  the relations 
between the magnitudes and the spectral type reported by
\citet{ruelas-mayorga:1991} and \citet{johnson:1966}.
Absolute magnitudes have been computed from
 the distances reported in Table \ref{table_sources}, and correction
of the reddening has been estimated from the absorption
measured either from X--ray or optical observations, based on the relation
$\Av \mbox{(mag)} = 5.59 \times 10^{-22} \nh (\cmmoinsdeux)$
\citep{predehl:1995}. 
The absorption  in the infrared  bands J, H and K is given by 
$\frac{A_{J}}{A_{V}} = 0.282$, 
   $\frac{A_{H}}{A_{V}} = 0.175$ and
   $\frac{A_{K}}{A_{V}} = 0.112$
\citep{rieke:1985}.
We report the absolute infrared magnitudes of our targets
on a CMD in Figure \ref{hr}.
We report in Figure \ref{asm} the X-ray lightcurves of the sources
as observed by the All Sky Monitor (ASM) of the 
Rossi X-ray Timing Explorer ({\it RXTE}) in the interval 
of our observations.

\subsection{GRO J1655-40}

                \subsubsection{Introduction}

$\gro$ (Nova Scorpii 1994) was discovered with BATSE 
as a hard X-ray Nova \citep{zhang:1994} and 
was the second superluminal source to be discovered in the Galaxy 
\citep{hjellming:1995a}. 
The column density was estimated to be in the range 
$3-8 \times 10^{21} \cmmoinsdeux$
(\citealt{hynes:1998}, \citealt{inoue:1994}, \citealt{greiner:1996a}, 
\citealt{inoue:1995}).
Following \citet{hynes:1998}, we adopt the intermediate ASCA measurement of 
$\nh = 4.4 \times 10^{21} \cmmoinsdeux$ \citep{nagase:1994}.
Assuming a mean extinction along the line of sight, the distance of the source
has been estimated as $3 \kpc$ \citep{greiner:1996a}.
$\gro$ has a bright variable 
optical counterpart \citep{bailyn:1995}: 
$B \sim [19-18.4]$, $V \sim [17.4-16.8]$, 
$R \sim [16.4-15.8]$ and $I \sim [15.4-14.8]$ (\citealt{orosz:1997}, 
\citealt{greene:2001}). 
Comparing the spectrum of the source in quiescence
with many standard stellar spectra of type M and K, 
the companion star was estimated 
to have a spectral type F3-F6. Following arguments on the size
of the star compared to its Roche lobe, it was argued that
the star was a sub-giant \citep{orosz:1997}.

The masses of the two components of the system have been precisely
determined thanks to optical observations: 
the compact object mass is in the range $4.1-7.9 \Msol$ (see 
\citealt{phillips:1999}, \citealt{shahbaz:1999} and \citealt{soria:1998}), 
making $\gro$ a
very good black hole candidate 
while the secondary star mass was estimated at $1.7 - 3.3 \Msol$,
both with 95 \% confidence \citep{shahbaz:1999}. 
The spectroscopic period is $2.62157 \pm 0.00015$ days \citep{orosz:1997}, 
with a radial velocity semi-amplitude of $K = 215.5 \pm 2.4 \kms$
and a mass function  $f(M)  = 2.73 \pm 0.09 \Msol$ \citep{shahbaz:1999}. 
The position of the secondary on the 
Hertzsprung-Russell diagram has been claimed to be 
consistent with a star of $\sim 2.3 \Msol$, which evolved from
the main sequence, and is now mid-way between the main sequence 
and the beginning of the giant branch \citep{kolb:1997}.

        \subsubsection{The observations}

As it has a bright optical counterpart, $\gro$ has not been studied in 
great detail in NIR. 
In Table \ref{table_obs} we present the only NIR
photometry which has been reported for this source, 
the counterpart being seen in the three filters.
Our 1997 observations were performed at the end of an X-ray outburst 
(see Figure \ref{asm}), consequently
the NIR emission is likely contaminated by an external
source.
We take therefore the magnitudes corresponding to the observations in an
almost quiescent state by \citet{greene:2001}.
The absolute magnitudes are $M_J = 0.77 \pm 0.15$ and $M_K = 0.59 \pm 0.15$,
the error quoted being due to the ellipsoidal variation.

We can see in Figure \ref{hr} that the colours and magnitudes of 
the source in quiescence locate it on the CMD between the
main sequence and giant star branches.
This is therefore consistent with the sub-giant luminosity class derived by
\citet{orosz:1997} and also with the fact that the source is 
crossing the Hertzsprung gap.
However, the position in the CMD 
shows a discrepancy with the F3-6 spectral type mentioned earlier.
Therefore, there is an emission which is not of stellar origin
in the NIR emission of $\gro$. Furthermore, this emission 
does not seem to be due
to irradiation since the ellipsoidal lightcurve of the 1999 observations
is well fitted by a model without any disk \citep{greene:2001}. 
This discrepancy was also pointed out by \citet{beer:2001}
in their recent analysis of its quiescent light curve.
Monitoring of the NIR emission during different
states of activity of this source will be necessary 
to reveal the origin of this emission.

\subsection{GRS 1739-278} \index{GRS 1739-278}

        \subsubsection{Introduction}

$\grsdstn$ is a hard X-ray transient source, discovered by SIGMA on 1996,
March 18th \citep{paul:1996}. 
The hardness of its spectrum immediately suggested that
it was an X-ray nova containing an accreting black
hole. $\grsdstn$ seems to be located near the galactic center, 
therefore at the distance of $\sim 8.5 \kpc$ \citep{marti:1997a}.
\citet{vargas:1997} inferred a peak luminosity of 
$8.6 \pm 2.0 \times 10^{36} \ergs$
in the 40-300 keV energy band.
A variable radio source in the
hard X-ray error box was proposed as the counterpart of $\grsdstn$
\citep{hjellming:1996}.
A candidate optical/infrared counterpart was soon discovered
at the position of the radio counterpart
with a constant luminosity in a range of $0.2 \mags$ 
on a timescale of several weeks during 1996 \citep{mirabel:1996d}. 
The observed optical magnitudes of $\grsdstn$ are $V = 23.2 \pm 0.3$, 
$R = 20.5 \pm 0.1$ and $I = 18.3 \pm 0.3$ 
\citep{marti:1997a}.
The magnitudes and colours of the companion star of $\grsdstn$ seemed 
to suggest either a low-mass X-ray binary 
with a giant companion, or a high-mass X-ray binary. 
The major problem in distinguishing between them
was the great uncertainty in the value of the hydrogen column density 
\citep{marti:1997a}.
The $\grsdstn$ column density estimates range from 
$1.2 \pm 0.1 \times 10^{22} \cmmoinsdeux$ \citep{greiner:1996b} to
 $4.1 \pm 0.7 \times 10^{22} \cmmoinsdeux$ \citep{borozdin:1996}.

        \subsubsection{The observations} \label{grsdstn_obs}

The counterpart is confirmed by our observations, showing 
a continuous decline in the luminosity of this source 
(see Table \ref{table_obs}). The source dropped by 
$\geq 3$ magnitudes in both J and K between 1996 and 1998.
For our analysis we will consider the magnitudes of 1998
as upper limits.
Taking the assumed distance of $D \sim 8.5 \kpc$, and 
the intermediate value of the column density 
$\nh = 2.0 \times 10^{22} \cmmoinsdeux$ \citep{marti:1997a}, 
we can derive the absolute magnitudes respectively in the J and K bands: 
$M_J \geq 1.40 \mbox{ and } M_K \geq 2.5$.
J-K is not constrained; this source lies below the line 
$M_K = 2.5$ on the CMD (Figure \ref{hr}). 
If $\grsdstn$'s companion star is on the main
sequence then it must be F5 V or later.
By examining the optical and near-infrared colours of the system, 
\citet{marti:1997a} derived two possibilities for the nature
of the secondary star: either a luminous early/middle B type main
sequence star, or a middle G/early K giant star.
Clearly, the implicit assumption by
\citet{marti:1997a}
that the source had reached the quiescent level was
premature at that time, and the magnitudes reported here 
allow us to better constrain the spectral type of this system.

\subsection{GRS 1716-249}

        \subsubsection{Introduction}

$\novoph$ (Nova Ophiuchi 1993) 
is an X-ray transient source, detected on 1993, 
September 25th by SIGMA on
 GRANAT, and by BATSE on 
the $\gamma$-ray observatory Compton
(GRO J1719-24) \citep{ballet:1993}. 
Its light curve during the flare was very similar to the one of $\novmus$,
and the (0.1-100 keV) X-ray luminosity at maximum was $L_{X} \sim 2.1
\times 10^{38} \ergpars$, which is close to the Eddington limit for a
compact object of $1.6 \Msol$. This X-ray luminosity is 
similar to those of A 0620-00 and of $\novmus$, both of which are, 
like $\novoph$,
transient radio sources.
ASCA observations gave an estimation of the column density of
$N_{H} = 4 \times 10^{21} \cmmoinsdeux$ \citep{tanaka:1993}.
The optical counterpart was soon discovered \citep{dellavalle:1994}. 
From $V= 16.65$ \citet{dellavalle:1994}
derived that the companion star was 
a low mass main sequence star of spectral type $\sim$ K or later.
This classification was consistent with 
the photometric and spectroscopic properties
of this object. 

The distance of this source remains subject to uncertainties.
Estimated from the equivalent width of the NaD absorption lines
the derived distance is $D \sim 2 \kpc$, while
taking the mean absolute magnitude at the maximum 
of the low mass X-ray binaries, 
the distance has been estimated to be $\sim 2.8 \kpc$,
giving an absolute magnitude $M_V \geq 6$ \citep{dellavalle:1994}.
\citet{masetti:1996} discovered a superhump period at 14.7 hours, therefore
indicative of the orbital period at a few percent accuracy. They estimated
the mass respectively of the primary and the companion star
to be $\geq 4.9 \Msol$ and $1.6 \Msol$. As noted by the authors,
the secondary would then be substantially brighter than
claimed by \citet{dellavalle:1994}, suggesting 
either the distance has been underestimated, or the secondary
is a slightly evolved late type star.

        \subsubsection{The observations}

$\novoph$ was not detected in our 1998 observations 
(see Table \ref{table_obs}), 
which were less sensitive that the 1997 ones in which 
the source was still visible at faint fluxes.
Other observations with more powerful telescopes are needed in order to see
if $\novoph$ has now reached its minimum, or if its luminosity is still
decreasing.
We will take for this analysis 
the magnitudes of 
1997, assuming that they correspond to a minimum.
Adopting a distance of
 $D \sim 2.4 \kpc$ and a column density
$N_{H} = 4 \times 10^{21}\cmmoinsdeux$, we can deduce
the absolute magnitudes respectively in the J, H and K bands: 
$M_J = 6.9 \pm 1.2 \mbox{ , } M_H = 6.9 \pm 1 \mbox{ and } M_K = 6.1 \pm 1$.
This allows us to say that the counterpart is a 
main sequence star of
spectral type M0-5 V. This is not consistent with the possible masses
of the companion star derived by \citet{masetti:1996}.
It seems likely therefore 
that the secondary is a slightly evolved late type star.
Therefore, our NIR absolute magnitudes are 
consistent with the absolute magnitude $M_V \geq 6$ of the optical counterpart 
identified by \citet{dellavalle:1994}, 
 and furthermore allow us to constrain better the nature of the companion star.

\subsection{GRS 1121-68} \index{Nova Muscae 1991}

        \subsubsection{Introduction}

This X-ray nova (Nova Muscae 1991) was discovered by Ginga on 1991 January 8th 
and by GRANAT on 1991 January 9th \citep{lund:1991}. 
The column density has been derived from ROSAT and Ginga observations:
$\nh = 2.2 \times 10^{21}\cmmoinsdeux$ \citep{greiner:1994}. 
The optical counterpart was identified by
\citet{dellavalle:1991} with a star which rose from $R \sim 20$ 
to $V \sim 13.3$.
The distance of this object has been subject to many uncertainties.
The estimation from E(B-V) gives $D = 2.3 \pm 2.1 \kpc$, but using the
linear relation between the equivalent width of the NaD line and the
distance, a distance of $D = 1.4 \kpc$ could be derived 
\citep{dellavalle:1991}.
A distance of $2.8 \kpc$ was estimated from observations in the H-band, with
an upper limit of 4 kpc \citep{shahbaz:1997}. 
Optical observations in quiescence, when its magnitude after dereddening was $B_0 = 19.8$,
showed that $\novmus$ is a low mass X-ray binary composed of a black hole
of mass greater than $\sim 3 \Msol$ and of a 
$0.7 \Msol$ low mass companion of
spectral type in the interval K0-4 V \citep{remillard:1992}.
Further optical spectroscopic observations in quiescence
suggested the donor star
to be a low main sequence star of spectral type of K3-5 V
 \citep{orosz:1996} and allowed the detection of a 10.5 hour orbital period
\citep{bailyn:1992}.

        \subsubsection{The observations}

To estimate the nature of the binary system, we selected from 
Table \ref{table_obs} the magnitudes
of the source near minimum luminosity, i.e. in 1994.
Following \citet{shahbaz:1997}, 
we choose 2.8 kpc as the most likely distance.
The absolute magnitudes estimated in the J and K bands are respectively: 
$M_J = 4.92 \pm 0.25 \mbox{ and } M_K = 4.33 \pm 0.25$.
We can see on Figure \ref{hr} that the location of this point
on the CMD is consistent with the companion star being a 
main sequence star of spectral type K0-5 V, so fully consistent
with the previous spectroscopic observations.
This shows that in the case where the NIR only comes from the
companion star, i.e. when there is no contamination,
this method can be efficiently used 
to constrain the spectral type of the companion star.

\subsection{GX 339-4} \index{GX 339-4}

        \subsubsection{Introduction}

GX 339-4 was discovered in 1973 by the 1-60 keV 
X-ray MIT detector on the satellite  OSO-7 \citep{markert:1973}. 
Because of its X-ray spectral behaviour similar to that of $\cygxu$
and of its rapid temporal
variations (from 0.010 to 10 s), $\gx$ was classified as a
black hole candidate \citep{tanaka:1996}.
The counterpart of $\gx$ was identified as a blue
star of typical $V = 16.6$ magnitudes but variable between $15<V<21$ 
magnitudes (\citealt{doxsey:1979}, \citealt{grindlay:1979}).
 A 14.8 hour modulation of the 
optical luminosity was interpreted as the orbital period of the
binary system \citep{callanan:1992}. 
However, because of the substantial optical emission
from the accretion disk, the orbital parameters of $\gx$ have not
yet been established in order to clearly demonstrate that it is a black hole
binary: the estimated mass of the compact object is
$\leq 2.5 M_{\odot}$ \citep{cowley:1986}.

Recent optical observations of $\gx$ in an extended ``off'' state allowed
to estimate a lower limit to the distance of $5.6 \kpc$ 
and an evolved spectral type later than F8 \citep{shahbaz:2001}.
\citet{zdziarski:1998} derived the extinction 
$E(B-V) = 1.2 \pm 0.1 \mags$ 
which is equivalent to $A_v = 3.72 \pm 0.1 \mags$ 
(see e.g. \citealt{cardelli:1989}).

        \subsubsection{The observations}

The results (Table \ref{table_obs}) show that the
luminosity changed appreciably during this period: 
by 2 magnitudes in J; 1.8 in H; and $\sim 2.6$ in K. 
The X-ray activity is shown in Figure \ref{asm}. We take the 
magnitudes corresponding to the minimum luminosity of this source
in 1993,
the two observations 
being taken with a one day interval.
At this date the source was not detected with BATSE:
the source was certainly in a low state (off state) or
a low low-hard state (S. Corbel, private communication).
Using the distance $D = 5.6 \kpc$
and an absorption of 3.72
magnitudes in the V-band,
we obtain the absolute magnitudes respectively
in the J and K-bands: $M_J = 1.41 \pm 0.3$ and $M_K = 1.04 \pm 0.3$.

From the location of $\gx$ on the CMD, 
it appears that the companion star is a red giant of 
spectral type F8-G2 III. 
This evolved type is consistent with the analysis of \citet{shahbaz:2001}
and with the orbital period of 14.8 hours
of the binary system.

\section{Discussion and general conclusions} \label{discussion}

NIR photometry is useful for constraining the stellar spectral
 type of the secondary star when the source
is heavily obscured at optical wavelengths.
We have determined constraints on the stellar spectral type 
of the counterparts of the galactic black hole candidates
$\gro$, $\grsdstn$, $\novoph$, $\novmus$ and $\gx$.
Our results are summarized in Table \ref{table_results}
and displayed in Figure \ref{hr}. 
All the sources but $\gro$ are consistent with low-mass
 stars as the companion star of the binary system.
The position of each source in this CMD
allows us to roughly estimate the evolutionary state of the secondary
while its J-K colour allows us to see if the infrared emission 
only comes from the photosphere of the companion or is 
contaminated by an external source.

The most important results are 
our constraints on the
companion stars in $\novoph$, $\novmus$ and $\gx$,
along with the weaker constraint on the companion of $\grsdstn$.
For the sources $\novoph$ and $\novmus$, 
the location in the CMD diagram is fully consistent with
their magnitudes, indicating that the infrared emission 
mainly emanates from the companion star, without the need for
any other source of emission. The derived spectral types are respectively
M0-5 V for $\novoph$, K0-5 V for $\novmus$ and F8-G2 III for $\gx$.
If the companion of $\grsdstn$ is on the main sequence then it
must be a low mass star of spectral type F5 V or later.
Concerning $\gro$, its location, between the main sequence
and giant star branches, is consistent with the sub-giant luminosity
class and with this source crossing the Hertzsprung gap.
However, there is a discrepancy with the optically determined
F3-6 spectral type, showing that a non-stellar emission seems to contribute 
to the NIR flux of this source.

\begin{table*}
\begin{flushleft}
\begin{tabular}{|c|c|c|c|c|c|c|c|}
\hline
Source     & 
             A$_J$            & A$_H$            & A$_K$            & 
             $M_J$            & $M_H$            & $M_K$            & 
             Spectral type \\
           & \multicolumn{6}{c|}{(mag)}                      &               \\ \hline
$\gro$     & 
             1.10          & 0.59          & 0.44         & 
             $0.77$        & -             & $0.59$       & 
             F3-6 IV       \\
           & 
                           &               &              &
             $\pm 0.15$    & -             & $\pm 0.15$   & 
                           \\ \hline
$\grsdstn$ & 
             3.15          & 1.96          & 1.25         & 
             $\geq 1.40$   & -             & $\geq 2.5$   & 
             $\geq$ F5 V   \\ \hline
$\novoph$    & 
             0.63          & 0.39          & 0.25         & 
             $6.9$         & $6.9$         & $6.1$        & 
             M0-5 V       \\ 
             &
                           &               &              &
             $\pm 1.2$     & $\pm 1$       & $\pm 1$      &
                           \\ \hline
$\novmus$    & 
             0.35          & 0.22          & 0.14         & 
             $4.92$        & -             & $4.33$   & 
             K0-5 V       \\ 
             &
                           &               &              &
             $\pm 0.25$    &               & $\pm 0.25$   &
                           \\ \hline
$\gx$        & 
             1.05          & 0.65          & 0.42         & 
             $1.41$        & -             & $1.04$       & 
             F8-G2 III \\ 
             &
                           &               &              &
             $\pm 0.3$     &               & $\pm 0.3$    &
             \\ \hline
\noalign{\smallskip}
\end{tabular}
\end{flushleft}
\caption[]{\label{table_results} Results corresponding to the minimum
of luminosity chosen during all our observations. We report here
the absorption, and the
corresponding derived absolute magnitudes.
We note that the spectral type of $\gro$ does not come from our
analysis, but from preceding optical observations. Only the sub-giant
luminosity class is consistent with the NIR observations.
}
\end{table*}

\section{acknowledgements}
S.C. is grateful to U. Kolb for stimulating discussions,
to J.A. Orosz for comments on $\gro$,
to C.A. Haswell for improving the language of the manuscript
and for useful suggestions, and finally to
S.J.B. Burnell for a careful rereading of the manuscript.
S.C. also acknowledges the anonymous referee for 
useful comments which allowed to improve the paper.
We acknowledge the quick-look results provided by the 
ASM/RXTE team, used to produce the Figure \ref{asm}.
S.C. acknowledges support from grant F/00-180/A from the Leverhulme Trust.
I.F.M. acknowledges support from CONICET/Argentina.
J.M. acknowledges partial support by DGICYT (PB97-0903)
and by Junta de Andaluc\'{\i}a (Spain).


\end{document}